\documentclass[11pt,article]{amsart}
\usepackage{amssymb}
\usepackage{amsfonts}
\usepackage{amsbsy,url}
\usepackage{latexsym}
\usepackage{amssymb,latexsym,amsmath,amsthm}
\usepackage[normalem]{ulem}
\usepackage{color}
\theoremstyle{plain}

 \theoremstyle{remark} 

\newcommand{\C}{{\mathbb C}}

\newcommand{\rank}{\operatorname{rank}(}

\newtheorem {theo} {\bf Theorem} [section]
\newtheorem {prop} [theo] {\bf Proposition}
\newtheorem {coro} [theo] {\bf Corollary}
\newtheorem {lem} [theo] {\bf Lemma}
\newtheorem {defi} {\bf Definition}[section]
\newtheorem{exam} {\bf Example}[section]

\newtheorem{rem}{\bf Remark}[section]

\numberwithin{equation}{section}
\begin{document}
\title{Phase retrievability of frames and quantum channels }


\author{Kai Liu, Deguang Han}
\date{\today}

%
\begin{abstract} 
A phase retrievable quantum channel  refers to a quantum channel $\Phi: B(H_A)\to B(H_B)$ such that there is a positive operator valued measure (POVM) $\{F_{j}\}$ in $B(H_{B})$ and $\{\Phi^*(F_j)\}$ is a phase retrievable operator valued frame. In this paper we examine the phase retrievable quantum channels in terms of  their  Kraus representations. For  quantum channels $\Phi$ of Choi's rank-$2$, we obtain a necessary and sufficient condition under which it is phase retrievable. For the general case, we present several necessary and/or sufficient conditions. In particular,  a necessary and sufficient condition is obtained in terms of the relevant  matrix-valued joint spectrum of the Kraus operators. Additionally, we also examine,  by examples,  the problem of constructing quantum channels such that there exists a minimal number of rank-one observables $\{F_{j}\}$ such that  $\{\Phi^*(F_j)\}$ does phase retrieval for $H_A$. Conversely, for a given set of rank-one observables $\{F_{j}\}_{j=1}^{N}$, we present a sufficient condition under which, for every $1\leq r\leq N$ given, a phase retrievable quantum channel $\Phi$ of Choi's rank-$r$ can be explicitly constructed.
\end{abstract}

\maketitle

\section{Introduction}


A quantum channel is a model for a particular snapshot of the time evolution of a density matrix, and especially
for the evolution from pure to mixed states. Let $H_A$ and $H_B$  be two quantum systems which are usually considered as finite dimensional Hilbert spaces.  A quantum channel $\Phi$ in  the Schr\"{o}dinger picture is  a  completely positive trace-preserving (CPTP for short) linear map between operator systems $B(H_A)$ and $B(H_B)$, and a quantum channel in the Heisenberg picture is described by the  adjoint map of a CPTP map (c.f.  \cite{Paulsen, 1Shankar, 2Wilde}).  By the famous theorem of Kraus (1971) (c.f. \cite{Choi1, 3Nielsen}), a  completely positive  map $\Phi$ from $B(H_A)$ to $B(K_B)$ has a  Kraus representation of the form:
$$
\Phi(T) = \sum_{i=1}^{r}A_iTA_i^*, \ \forall T \in B(H_A)
$$
for some operators $A_1, . . . , A_r\in B(H_A, H_B)$. In this representation,  $A_1, ... , A_r$ are also referred as the Kraus operators of $\Phi$. It is easy to see that such a map $\Phi$ is trace-preserving  if and only if $\sum_{i=1}^{r}A_i^*A_i = I_{H_A},$  and it is unital if and only if $\sum_{i=1}^{r}A_{i}A_{i}^* = I_{H_B}$  (c.f. \cite{4Gupta}).

 Recall that a positive operator valued measure (POVM for short) or observables on a Hilbert space $H$  is a collection of positive  operators $\{F_i\}$ in $B(H)$ such that $\sum_{j\in \Bbb{J}} F_j = I_H$. A POVM $\{F_{j}\}_{j\in\Bbb{J}}$ is  {\it information complete} (c.f.  \cite{5Renes}) if $\{\langle x, F_jx\rangle\}_{j\in \Bbb{J}}$ uniquely determines  the pure state $\rho_x : = x\otimes x$ (the rank-one projection determined by $x$). In other words, an information complete POVM is a phase retrievable operator valued frame. 
  For a quantum channel $\Phi: B(H_A)\to B(H_B)$,  its adjoint $\Phi^*$ is unital and hence $\{\Phi^*(F_{j})\}_{j\in\Bbb{J}}$ is a POVM for $H_A$ whenever $\{F_{j}\}_{j\in\Bbb{J}}$ is an POVM for $H_B$. In the Heisenberg picture, a POVM in $H_B$ are the observables that are used to measure a state $\rho$ in $B(H_A)$ with measurement $\langle \rho, \Phi^*(F_j)\rangle = tr(\rho\Phi^*(F_j)) = tr(\Phi(\rho)F_j)$. It is important that a quantum channel $\Phi$ admits a POVM  on $H_B$  that distinguishes the pure states from $H_A$ (c.f. \cite{7Ariano, 6Tumalka}). For this reason, we formally introduce  the following definition:
   
 \begin{defi} A quantum channel $\Phi: B(H_A)\to B(H_B)$ is called {\it phase retrievable} if there exists a  POVM $\{F_{j}\}_{j\in\Bbb{J}}$ in $B(H_B)$ such that $\{\Phi^*(F_{j})\}_{j\in\Bbb{J}}$ is a phase retrievable operator valued frame for $H_A$.
 \end{defi}

The main purpose of this paper is to characterize the phase retrievable  quantum channels in terms of their Kraus operators. We first recall some necessary background materials that are needed in this paper. 

Notations:  Let $H$ and $K$ be separable complex Hilbert spaces.
\begin{itemize}

\item $B(H, K)$ -- the space of bounded linear operators from $H$ to $K$, and write $B(H) = B(H, H)$.
\item $B_{sa}(H)$ -- the set of all self adjoint operators in $B(H)$.
\item $\langle S, T\rangle = tr(ST^*)$ -- the Hilbert-Schmidt inner product on the space of trace-class operators

\item $\rho_{x, y} = x\otimes y$ --  the rank-one operator defined by $(x\otimes y)(z) = \langle z, y\rangle x$ for every $z\in H$, and write $\rho_x = \rho_{x, x}$.

\item $M_{n\times n}(B(H))$ -- the matrix algebra over $B(H)$, i.e., the set of all the $n\times n$ matrices with entries in $B(H)$.
\end{itemize}

\subsection{Phase Retrievable Frames} Frame theory has close connections with operator valued measures and consequently with quantum information theory. In fact, it offers 
a lens through which to view a large portion of quantum information theory,  and to tackle problems such as quantum detections, information completeness, channel designs etc. We refer to \cite{Beneddeto}-\cite{BCCHT1}, \cite{7Ariano}-\cite{Han-5} and \cite{Poumai}-\cite{5Renes} for these connections and some recent developments related to this topic. The relevant part to the frame theory in this paper is phase retrievable frames. 
The phase retrieval problem, i.e., the problem of recovering a signal $x$ in a Hilbert space $H$ up to a unimodular scalar from a system of measurements, which appears in many applications. This is the same as recovering the rank-one operator $x\otimes x$ with a given measurement. While the phaseless measurements have been performed in various of schemes, lots of recent  research work have been focused on the frame-based approach particularly in the finite dimensional settings (c.f. \cite{14Akrami}-\cite{10Bodmann}, \cite{CEHV-ACHA, Edidin} \cite{Eldar}-\cite{9Eldar} \cite{Han-2, 11Li} and \cite{Wang-Xu-1}-\cite{Wang-Xu-2}).

A sequence $\{f_j\}_{j\in \Bbb{J}}$ is called a {\it frame}  for a Hilbert space $H$ if there are two positive constant numbers  $A,B>0$ such that $$A||x||^2\leq\sum_{i\in I}|\langle x, f_i\rangle |^2\leq B||x||^2$$ holds for every $x\in H$. A frame is called a {\it tight frame} if $A= B$ and a {\it Parseval frame} if  $A= B =1$. A frame  $\{f_j\}_{j\in \Bbb{J}}$  is a Parseval frame if and only if $\sum_{j\in \Bbb{J}} f_j\otimes f_j = I$. Every frame  $\{f_j\}_{j\in \Bbb{J}}$ is similar to a Parseval frame in the sense that there is an invertible operator $S\in B(H)$ such that $\{Sf_j\}_{j\in\Bbb{J}}$ is a Parseval frame (c.f.  \cite{13Han}).

A frame $\{f_j\}_{j\in \Bbb{J}}$ is called {\it phase retrievable} if the magnitudes of its frame coefficients $\langle x, f_j\rangle $ of a  signal $x\in H$ uniquely determines $\rho_x$. More generally, a collection of operators $\{A_j\}_{\in\Bbb{J}}$ in $B(H)$ is called a {\it phase retrievable operator valued frame}  for $H$ if the phaseless measurements $\langle A_jx, x \rangle = \langle \rho_x, A_j\rangle$ uniquely determines $\rho_x$. It is obvious that a (vector-valued) frame $\{f_j\}_{j\in \Bbb{J}}$ is phase retrievable if and only if $\{f_j\otimes f_j\}_{j\in \Bbb{J}}$ is a phase retrievable  operator valued frame. Phase retrievable frames are closely related to the concept of complement property: A frame $\{f_j\}_{j\in \Bbb{J}}$ for $H$ is said to have  the complement property if for every $\Omega\subset \Bbb{J}$, we have either $\overline{span}\{f_i\}_{i\in \Omega}=H$ or $\overline{span}\{f_j\}_{j\in \Omega^c}=H$. 
The following is well-known. 

\begin{prop} \label{prop-1.1} Let  $H$ be an $n$-dimensional Hilbert space over $\Bbb{F} = \Bbb{R}$  or $\Bbb{C}$, and let $\mathcal{F}=\{f_j\}_{j=1}^{N}$ be a frame for $H$. Then the following statements are true.

(i)  $\{f_j\}_{j=1}^{N}$ has the complement property if and only if $span\{\langle x,  f_j\rangle f_j: j =1, ... ,N\} = H$ for every $x\neq 0$.

(ii) If $ \{f_i\}_{j\in \Bbb{J}}$ is phase retrievable, then  it has the complement property. The complement property is also  sufficient when $\Bbb{F} = \Bbb{R}$.

(iii)  If $\Bbb{F} = \Bbb{R},$  then every generic frame $\{f_j\}_{j=1}^{N}$ of length $N\ge 2n-1$ is phase retrievable. If $\Bbb{F} = \Bbb{C},$  then every generic frame $\{f_j\}_{j=1}^{N}$ of length $N\ge 4n-4$ is phase retrievable. 

(vi) If  $n = 2^{k} +1$ and $\{f_j\}_{j=1}^{N}$ is phase retrievable for $C^n$, then $N \geq 4n-4.$

(v) Let $L_{\mathcal{F}}: B(H)\to \Bbb{F}^N$ be the linear map defined by $L_{\mathcal{F}}(T) = (tr(Tf_j\otimes f_j))_{j=1}^{N}$, and $\mathcal{S}_{2}$ be the set of all the Hermitian operators with rank less than or equal to $2$. Then $\{f_{i}\}_{i=1}^{N}$ is phase retrievable if and only if $ker L_{\mathcal F}\cap \mathcal{S}_{2} = \{0\}.$
\end{prop}

Note that the  complement property implies that $2n-1$ is the minimal number of measurements needed to do phase retrieval in the real Hilbert space case, the question of finding the exact minimum number of measurements necessary to satisfy phase retrieval  in the complex case remains open (c.f. \cite{17Balan}).

More generally, a collection of self-adjoint operators $\{M_j\}_{\in\Bbb{J}}$ in $B(H)$ is called a  {\it phase retrievable operator valued frame} for $H$ if the measurements $\{\langle \rho_x, M_j\rangle\}_{j\in\Bbb{J}}$ uniquely determines $\rho_x\in B(H)$. Similar to Proposition \ref{prop-1.1} we have the following:

\begin{prop}  \cite{8Han, Wang-Xu-1} \label{prop-1.2} Let $\{M_j\}_{j\in\Bbb{J}}$  be a collection of self-adjoint operators in $B(H)$. If $\{M_j\}_{j\in\Bbb{J}}$  is phase retrievable, then
$$
span\{M_jx: j\in\Bbb{J}\} = H
$$
for every nonzero vector $x\in H$. This condition is also sufficient when $H$ is a real Hilbert space.
\end{prop}

\subsection{Kraus Representations  and Choi's Theorems}

Let $\Phi: B(H_A)\to B(H_B)$ be a linear map, $\{e_i\}_{i=1}^{n}$ be a basis of $H$ and $E_{ij}$ be the rank-one operator $e_i\otimes e_j$. Then the Choi's matrix $C_{\Phi}$ is the matrix  defined by
$$
C_{\Phi} = [ \Phi(E_{ij})]\in M_{n\times n}(B(H_A)),
$$
where $n = \dim H_A$.
By Kraus' theorem, every completely positive map $\Phi$ has a representation $\Phi(T) = \sum_{j=1}^{r}A_{j}TA_{j}^{*}$  for some $A_{j}\in B(H_A, K_B)$ and this representation is not unique.  The  {\it Choi rank} of  $\Phi$,  denoted by  $Cr(\Phi)$, is the smallest integer $r$ from the Kraus representations. The following Choi's Theorem is well known (c.f. \cite{4Gupta,17Paulsen,18Holevo}).

\begin{theo}  Let $\Phi: B(H_A)\to B(H_B)$ be a linear map. 

(i) Then $\Phi$ is completely positive (CP for short) if and only if  $C_{\Phi}$ is positive and further $Cr(\Phi) = \rank C_{\Phi})$ if $\Phi$ is a CP map.

(ii) Suppose that $\Phi(T) = \sum_{j=1}^{r}A_{j}TA_{j}^{*}$ with  $r = Cr(\Phi) $.    Then $\Phi(T) = \sum_{j=1}^{m}B_{j}TB_{j}^{*}$ if and only if   there exists matrix $U = (u_{ij})\in M_{m\times r}(\C)$ such that $U^{*}U  = I_{r}$ and $A_{i}  = \sum_{j=1}^{r}u_{ij}B_{j}$ for $i=1,... , m$
and $span\{B_{1}, ... , B_{m}\} = span\{A_{1}, ... , A_{r}\}$.
\end{theo}

\subsection{Pure-State Injectivity} 

A quantum channel $\Phi: B(H_A)\to B(H_B)$ is called {\it pure-state injective} if $\Phi(\rho_x) = \Phi(\rho_y)$ implies that $\rho_x = \rho_y$. The following observation tells us that pure-state injective quantum channels are exactly the phase retrievable ones in the Heisenberg picture.

 \begin{prop}\label{prop-1.3} A quantum channel $\Phi: B(H_A)\to B(H_B)$ is phase retrievable if and only if  $\Phi$ is pure-state injective.
  \end{prop} 
  \begin{proof} Suppose that $\Phi$ is phase retrievable, and let  $\{F_{j}\}_{j\in\Bbb{J}}$ be a POVM in $B(H_B)$ such that $\{\Phi^*(F_{j})\}_{j\in\Bbb{J}}$ is a phase retrievable operator valued frame for $H_A$. If $\Phi(\rho_x) = \Phi(\rho_y)$, then we have 
$$
\langle \rho_x, \Phi^*(F_j)\rangle = \langle \Phi(\rho_x), F_j \rangle =  \langle \Phi(\rho_y), F_j \rangle = \langle \rho_y, \Phi^*(F_j)\rangle
$$
for all $j\in\Bbb{J}$, and hence $\rho_x = \rho_y$. Therefore $\Phi$ is pure-state injective. Conversely, assume that  $\Phi$ is pure-state injective. Let $\{F_j\}_{j\in \Bbb{J}}$ be a POVM such that $span \{F_j: j\in\Bbb{J}\}$ coincides with the space of all self-adjoint operators. Then $\langle \rho_x, \Phi^*(F_j) = \langle \rho_y, \Phi^*(F_j)$ implies $ tr(\Phi(\rho_x)F_j) =  tr(\Phi(\rho_y)F_j)$ for all $j$, which implies that  $\Phi(\rho_x) = \Phi(\rho_y)$. Therefore $\rho_x = \rho_y$ and hence $\Phi$ is phase retrievable.
 \end{proof}
 
 The following example shows that pure-state injective quantum channel is not necessarily injective over $B(H)$. 

\begin{exam} \label{noninjective}

 By Proposition \ref{prop-1.1}, there exists a phase retrievable frame  $\{f_j\}_{j=1}^{N}$ for $H= \Bbb{C}^n$ such that $N < n^2$. We can also assume that $\{f_j\otimes f_j\}_{j=1}^{N}$ is linearly independent. 
Write $\rho_j = f_j\otimes f_j$, then $S = \sum_{j=1}^{N}\rho_j^2$ is positive invertible. Let 
$$
\Phi(T) = \sum_{j=1}^{N}\rho_j T\rho_j, \ \ T\in B(H).
$$
 Then $\Phi$ is pure-state injective but not injective on $B(H)$, so is  the quantum channel $$
\tilde{\Phi}(T) = \sum_{j=1}^{N}A_j TA^*_j, \ \ T\in B(H),
$$
where $A_j = S^{-1/2}\rho_j$.
\end{exam}
\begin{proof} 
Suppose that $\Phi(\rho_x) = \Phi(\rho_y)$ for some $x, y\in H$. Then we get
$$
 \sum_{j=1}^{N} |\langle x, f_j\rangle |^2 \rho_j =  \sum_{j=1}^{N} |\langle y, f_j\rangle |^2 \rho_j,
 $$
 which implies that $ |\langle x, f_j\rangle |^2  =  |\langle y, f_j\rangle |^2$ for all $j$. Thus $\rho_x = \rho_y$ since $\{f_j\}_{j=1}^{N}$ is phase retrievable. Therefore $\Phi$ is pure-state injective. However, with the same argument,  we have that $\Phi(T) = 0$ if and only if $\langle T, \rho_j\rangle = 0$ for all $j$. Since $N < n^2$ we know that $span\{\rho_j\} \neq B(H)$, and thus $\Phi$ is not injective.
\end{proof}

The main purpose of this paper is to characterize the phase retrievable quantum channels in terms of their Kraus operators. As a motivating case, we first examine the Choi's rank-2 quantum channels in section 2 and then the more general case. The conditions which make quantum channels to be phase retrievable are given in the form of  relative matrix-valued  joint spectrum of operator tuples. In section 3, we consider the problem of the existence and construction of phase retrievable quantum channels with minimal number of observables and prescribed Choi's rank. We will show that  such quantum channels can be constructed by using phase retrievable frame of the smallest length. 
  

\section{Characterization of phase retrievable quantum channels}

If a quantum channel $\Phi: B(H_A) \to B(H_B)$ has Choi rank-one, then $\Phi(T) = VTV^*$ for some $V\in B(H_A, H_B)$ such that $V^*V = I_{H_A}$. Clearly $\Phi$ is pure-state injective and hence it is always phase retrievable. However, the situation is much more complicated when $Cr(\Phi)$ is greater than one.  In this section, we present several necessary and/or sufficient conditions on the Kraus operators $\{A_1, ... , A_r\}$ under which $\Phi$ is phase retrievable. Such quantum channels will be characterized in terms of the {\it relative left joint spectrum} for a $k$-tuple of operators. For the sake of clarity, we will first deal with the case of $Cr(\Phi) =2$ and then discuss the general case.

\subsection{The Case of $Cr(\Phi) = 2$}

We need the following technical lemma:

\begin{lem}  \label{lem-key} Suppose that
	A=$\begin{pmatrix}
		a_1&b_1\\
		a_2&b_2
	\end{pmatrix} \in M_{2\times 2}(\C)$  satisfies the conditions
	\begin{equation}
	\begin{aligned}
		1+|a_1|^2-|a_2|^2=0, \\
		1+|b_2|^2-|b_1|^2 =0, \\
		a_1\bar{b}_1=a_2\bar{b}_2.
	\end{aligned}
	\end{equation}
 Then  $A$ has two  distinct eigenvalues $\lambda_1$ and $\lambda_2$, and  $\lambda_1\bar{\lambda}_2=-1$.
\end{lem}
\begin{proof} Clearly, $a_2$ and $b_1$ are nonzero numbers.
Let $\lambda = \frac{a_1}{a_2}=\frac{\bar{b}_2}{\bar{b}_1}$. Then from (2.1), we have $|a_2|^2=\frac{1}{1-|\lambda|^2}$ and $|b_1|^2=\frac{1}{1-|\lambda|^2}$, which shows that $|\lambda|<1$ and $|a_2|^2=|b_1|^2\neq 0$. Write $b_1=\mu a_2$ with $\mu=e^{i\theta}\in \mathbb{C}$. Then we have 
\begin{equation}
	A=\begin{pmatrix}
		\lambda a_2&\mu a_2\\
		a_2&\bar{\lambda}\mu a_2
	\end{pmatrix}
	=a_2\begin{pmatrix}
		\lambda &\mu \\
		1&\bar{\lambda}\mu 
	\end{pmatrix}\triangleq a_2B,
\end{equation}
where $B = \begin{pmatrix}
		\lambda &\mu \\
		1&\bar{\lambda}\mu 
	\end{pmatrix}.
	$
Since $p(t)=det(tI-B)=t^2-(\lambda+\bar{\lambda}\mu)t+(|\lambda|^2-1)\mu$, we get that  $t_{1,2}=\frac{1}{2}[(\lambda+\bar{\lambda}\mu)\pm \xi^{\frac{1}{2}}]$ are the two eigenvalues of $B$,
where $ \xi=(\lambda+\bar{\lambda}\mu)^2-4(|\lambda|^2-1)\mu$. Note that
\begin{equation}
	\begin{aligned}
		\xi&=(\lambda+\bar{\lambda}\mu)^2-4(|\lambda|^2-1)\mu\\
		&=(\lambda^2+2|\lambda|^2\mu+\bar{\lambda}^2\mu^2)-4|\lambda|^2\mu+4\mu\\
		&=(\lambda^2\bar{\mu}+\bar{\lambda}^2\mu+4-2|\lambda|^2)\mu\\
		&\triangleq \delta\mu,
	\end{aligned}
\end{equation}
where $\delta := (\lambda^2\bar{\mu}+\bar{\lambda}^2\mu+4-2|\lambda|^2)\mu \geq 4$ since the real number $ \lambda^2\bar{\mu}+\bar{\lambda}^2\mu \leq 2|\lambda|^2$.
This implies that $\xi^{\frac{1}{2}}=\sqrt{\delta}e^{\frac{i\theta}{2}}$. Thus  $2t_1=(\lambda+\bar{\lambda}\mu)+\sqrt{\delta}e^{\frac{i\theta}{2}}$ and 
		$2t_2=(\lambda+\bar{\lambda}\mu)-\sqrt{\delta}e^{\frac{i\theta}{2}}$. So we have  $4t_1\bar{t_2}=2|\lambda|^2-(4-2|\lambda|^2)
	=4|\lambda|^2-4$ and hence $t_1\bar{t_2}=|\lambda|^2-1$. Therefore the two eigenvalues $\lambda_1= a_2t_1$ and $\lambda_2 = a_2t_2$ of $A$ satisfy the claim: $$\lambda_1 \bar{\lambda_2}=|a_2|^2t_1\bar{t_2}= \frac{1}{1-|\lambda|^2}(|\lambda|^2-1)=-1.$$
\end{proof}

\begin{theo} \label{main-thm-1}Suppose that $\Phi: B(H_A) \to B(H_B)$ is a quantum channel with Kraus operators $\{A_1, A_2\}$. Then $\Phi$ is phase retrievable if and only if for every $ \lambda \in \mathbb{C}$, either $A_1+\lambda A_2$ or $-\bar{\lambda}A_1+A_2$ is injective.
\end{theo}
\begin{proof}
First, suppose the quantum channel $\Phi$ is phase retrievable.  Then there exists a POVM $\{F_j\}_{j\in \Bbb{J}}$ in $B(H_B)$ such that $\{{\Phi^*(F_j)}\}_{j\in\Bbb{J}}$ is a phase retrievable operator valued frame. Assume  to the contrary that neither $A_1+\lambda A_2$ nor $-\bar{\lambda}A_1+A_2$ is injective for some $\lambda\in \mathbb{C}$. Then there exist nonzero vectors $x, y\in H_A$  such that ${(A_1+\lambda A_2)x=0}$ and ${(-\bar{\lambda} A_1+A_2)y=0}$. Thus for any $j\in\Bbb{J}$, we have 
	\begin{equation}
	   \begin{aligned}
		\langle x,\Phi^*(F_j)y\rangle&=\langle x,A_1^*F_jA_1y\rangle + \langle x,A_2^*F_jA_2y\rangle \\
		&=\langle A_1x,F_jA_1y\rangle+ \langle A_2x,F_jA_2y\rangle \\
		&=\langle -\lambda A_2x,F_jA_1y\rangle + \langle A_2x,\bar{\lambda} A_1y\rangle \\
		&=-\lambda \langle A_2x,F_jA_1y\rangle +\lambda \langle A_2x, F_jA_1y\rangle \\
		&=0,
	   \end{aligned}
	\end{equation}
 which implies $span_{j\in\Bbb{J}}\{\Phi^*(F_j)y\}\neq H_B$. This implies by Proposition \ref{prop-1.2} that $\{{\Phi^*(F_j)}\}_{j\in\Bbb{J}}$ is not a  phase retrievable operator valued frame. Therefore we have either $A_1+\lambda A_2$ or $-\bar{\lambda}A_1+A_2$ is injective for every $ \lambda \in \mathbb{C}$.

Conversely, suppose that for $\forall \lambda \in \mathbb{C}$ either $A_1+\lambda A_2$ or $-\bar{\lambda}A_1+A_2$ is injective. In particular for $\lambda=0$, we have either $A_1$ or $A_2$ is injective. Without losing of generality, we can assume $A_1$ is injective and hence $A_1^*A_1$ is invertible. Let   $f_i\in H_B (i=1, ... , M)$ be such vectors that $span\{f_i\otimes f_i\}_{i=1}^{M}\supseteq B_{sa}(H_B)$ and $\{f_i\otimes f_i\}_{i=1}^M$ is a POVM. We claim that $\{\Phi^*(f_i\otimes f_i)\}_{i=1}^M$ is a phase retrievable operator valued frame for $H_A$.

Assume to the contrary that $\{\Phi^*(f_i\otimes f_i)\}_{i=1}^M$ is not phase retrievable. Then there exist two vectors $x, y\in H_A$ linear independent such that for $\forall 1\leq i\leq M$, we have
$$\langle\Phi^*(f_i\otimes f_i),x\otimes x\rangle = \langle\Phi^*(f_i\otimes f_i),y\otimes y\rangle, $$ i.e., 
$$
		\langle A_1x\otimes A_1x+A_2x\otimes A_2x, f_i\otimes f_i\rangle
		=\langle A_1y\otimes A_1y+A_2y\otimes A_2y, f_i\otimes f_i\rangle.
$$
Thus $$A_1x\otimes A_1x+A_2x\otimes A_2x =A_1y\otimes A_1y+A_2y\otimes A_2y$$ since $span\{f_i\otimes f_i\}_{i=1}^{M}\supseteq B_{sa}(H_B)$.
Let $T = A_1x\otimes A_1x+A_2x\otimes A_2x$. 

If $rank(T)=1$, then $A_1x, A_2x$ must be linear dependent. Thus there exist $\lambda, \mu \in \mathbb{C}$ such that $A_2x=\lambda A_1x$ and $A_2y=\lambda A_1y$.
So we have  $(1+|\lambda|^2)A_1x\otimes A_1x=(1+|\mu|^2)A_1y\otimes A_1y$, which implies that $A_1x,A_1y$ are linear dependent.  Since $A_1$ is injective, we get that $x$ and $y$ are linearly independent, which leads to a contradiction. Therefore we must have $rank(T)=2$, and hence $A_1x, A_2x$ are linear independent.  This implies that $Range(T)=span\{A_1x, A_2x\}=span\{A_1y, A_2y\}=span\{A_1x, A_1y\}$. Thus there exist $a_1, a_2, b_1, b_2\in\mathbb{C}$ such that 
\begin{equation}
	\begin{aligned}
		A_2x&=a_1A_1x+b_1A_1y\\
		A_2y&=a_2A_1x+b_2A_1y.
	\end{aligned}
\end{equation}
This implies that
\begin{equation}
	\begin{aligned}
		A_1x\otimes[&(1+|a_1|^2-|a_2|^2)A_1x+(a_1\bar{b_1}-a_2\bar{b_2})A_1y]\\
		&=A_1y\otimes[(1+|b_2|^2-|b_1|^2)A_1y+(b_2\bar{a_2}-b_1\bar{a_1})A_1x],
	\end{aligned}
\end{equation}
and hence we get
\begin{equation}
	\begin{aligned}
		1+|a_1|^2-|a_2|^2=0\\
		1+|b_2|^2-|b_1|^2 = 0\\
		a_1\bar{b_1}=a_2\bar{b_2}.
	\end{aligned}
\end{equation}
Applying $P=(A_1^*A_1)^{-1}A_1^*$  to (2.5), we get that $PA_2x=a_1x+b_1y$ and $PA_2y=a_2x+b_2y$. This shows that $M=span\{x,y\}$ is invariant under $PA_2$. Let $S = PA_2|_{M}$. Then $S$ has a representation matrix 
$$
A= \begin{pmatrix}
		a_1&b_1\\
		a_2&b_2
	\end{pmatrix} 
$$ that satisfies the conditions of Lemma \ref{lem-key}. Therefore we have $spec\{S\}=\{\lambda_1, \lambda_2\}$ and $\lambda_1\bar{\lambda_2}=-1$.

We claim that for $\lambda = - \bar{\lambda_2}$, neither   $A_1+\lambda A_2$ nor $-\bar{\lambda}A_1+A_2$ is injective,  which will lead to a contraction to the assumption.
Indeed, let $u,v \in M$ be two linear independent vectors such that
$Su=\lambda_1 u$ and $Sv=\lambda_2 v$. Note that if $z=A_1w \in Range(A_1)$ for some $w\in H_A$, we must have that $A_1Pz=A_1(A_1^*A_1)^{-1}A_1^*A_1w=A_1w=z$. 

Since  $u, v \in span\{x, y\}$ and $A_2x,A_2y\in Range(A_1)$ by (2.5), we get  that $A_2u,A_2v\in Range(A_1)$ and consequently we have $$\lambda_1A_1u = A_1Su = A_1PA_2u=A_2u$$ and $$\lambda_2A_1v = A_1Sv =A_1PA_2v=A_2v.$$
That implies that  $(A_1+ \bar{\lambda_2}A_2)u=0$ and $(-\lambda_2 A_1 + A_2)v=0$ since $\lambda_1\bar{\lambda_2}=-1$, which completes the proof.
\end{proof}

\subsection{The general case}

For the general case, we need to introduce the concept of (matrix-valued) relative left joint spectrum of $k$-tuple of operators with respect to another $m$-tuple of operators. 
This is a natural generalization of the left joint spectrum of a $k$-tuple of operators $\overrightarrow{A}= (A_1, ... , A_k)^{t}$ in $B(H,  K)$, where ``t" denotes the transpose. Recall that $\overrightarrow{A}$ is  called {\it left invertible} if there exists a $k$-tuple $\overrightarrow{B} = (B_1, ... , B_k)^t$  in  $B(H, K)$ such that
$$
\sum_{j=1}^{k} B^*_jA_j = I.
$$
The {\it  left joint spectrum} of $\overrightarrow{A}$ is defined by
$$
\sigma_{\ell}(\overrightarrow{A}) = \{ \overrightarrow{\lambda} \in \C^k: \overrightarrow{A} -  \overrightarrow{\lambda} I \ \text{is not left invertible}\},
$$
where $\overrightarrow{\lambda} I= (\lambda_1 I, ... , \lambda_k I)^t$.

\begin{defi} Let $\overrightarrow{A}= (A_1, ... , A_k)^t$ be a $k$-tuple  and $\overrightarrow{B}= (B_1, ... , B_m)^t$ be a $m$-tuple of operators in $B(H,  K)$.  Then the  {\it left $ \overrightarrow{B}$-relative  joint spectrum } of $\overrightarrow{A}$  is defined to be the set of  all the matrices $\Lambda \in M_{m\times k}(\C) $ such that  $\overrightarrow{A}-\Lambda\overrightarrow{B}$ is not left invertible, where $\Lambda\overrightarrow{B} = (\sum_{i=1}^{m}\lambda_{i1}B_i, ...,  \sum_{i=1}^{m}\lambda_{ik}B_i)^t$. We use $\sigma_{\ell}(\overrightarrow{A},\overrightarrow{B})$ to denote the set of all such matrices $\Lambda$. 
\end{defi}

\begin{rem} \label{lem-spectrum} In the case that $m=1$ and $\overrightarrow{B} = (I)$, we immediately get  $\sigma_{\ell}(\overrightarrow{A},\overrightarrow{B}) = \sigma_{\ell}(\overrightarrow{A})$. Thus the above concept naturally generalizes  the concept of the left joint spectrum of $\overrightarrow{A}$.
\end{rem}

\begin{lem} Let  $\overrightarrow{A}, \overrightarrow{B}$ and $\Lambda$ chosen as above. Then $\Lambda \in \sigma_{\ell}(\overrightarrow{A},\overrightarrow{B})$ if and only if there exists nonzero vector $x$ such that  $A_jx = \sum_{i=1}^{m}\lambda_{ij}B_i x$ for every $j=1, ... , k$.
\end{lem}
\begin{proof} This follows from a well-known fact that a $k$-tuple $\overrightarrow{C} = (C_1, ... , C_k)^t$ is not left invertible if and only if there exists a nonzero vector $x$ such that $C_jx = 0$ for every $j=1, ... , k$, or equivalently $\overrightarrow{C}$ is left invertible if and only if $\sum_{j=1}^{k}C^*_jC_j$ is invertible. We include a short argument for self-completeness. Clearly, $\overrightarrow{C}$ is left invertible if $\sum_{j=1}^{k}C^*_jC_j$ is invertible. Conversely, assume that $\overrightarrow{C}$ is left invertible. Then there exist operators $D_j$ such that $\sum_{j=1}^{k}D_j^*C_j = I$. We claim that $\sum_{j=1}^{k}C^*_jC_j$ is invertible. If not, then there is a nonzero vector $x$ such that $\sum_{j=1}^{k}C^*_jC_j x =0$, which implies that $\sum_{j=1}^{k}||C_jx||^2 = \langle \sum_{j=1}^{k}C^*_jC_jx , x\rangle = 0$. Thus $C_jx= 0$ for each $j$ which contradicts with  $\sum_{j=1}^{k}D_j^*C_j = I$.
\end{proof}

The characterization of phase retrievable quantum channels are different for real and complex Hilbert spaces. Therefore we treat them separately in two subsequent subsections. 

\subsubsection{The real Hilbert space case} In this subsection we assume that $H_A$ and $H_B$ are both finite-dimensional real Hilbert spaces.

\begin{lem} \label{lem-2.1} Let $\Phi: B(H_A)\to B(H_B)$ be a real quantum channel. Then the following are equivalent:

(i) $\Phi$ is phase retrievable.

(ii) There exists a POVM $\{F_j\}_{j\in \Bbb{J}}$ such that $span_{j\in \Bbb{J}}\{\Phi^*(F_j)x\} = H_A$ for every nonzero vector $x \in H_A$.

(iii) $\Phi(x\otimes y) \neq 0$ whenever $x\otimes y \neq 0$.

Moreover, $(i) \Rightarrow (ii) \Leftrightarrow (iii)$ is also true for complex Hilbert space case. 
\end{lem}
\begin{proof} The equivalence of $(i)$ and $(ii)$ follows from the definition of phase retrievable quantum channel and Proposition \ref{prop-1.2}. So we only need to show that $(ii)$ and $(iii)$ are equivalent.

 Assume (ii) and  let $x, y\in H_A$ be any two nonzero vectors. Then $\forall j\in\Bbb{J}$, we have
\begin{equation}
\begin{aligned}
\langle \Phi(x\otimes y), F_j\rangle &=\langle x\otimes y, \Phi^*(F_j)\rangle=tr(x\otimes y(\Phi^*(F_j^*)))\\
&=tr((\Phi^*(F_j^*))(x\otimes y))=\langle (\Phi^*(F_j^*))x, y\rangle \\
&=\langle x,\Phi^*(F_j)y\rangle.
\end{aligned}
\end{equation} 
Since $y\neq 0$ we have  $span_{j\in \Bbb{J}}\{\Phi^*(F_j)y\}=H_A$. Thus there must exist $j_0\in \Bbb{J}$ such that $$\langle\Phi(x\otimes y), F_{i_0}\rangle = \langle x,\Phi^*(F_{j_0})y\rangle\neq 0,$$
which implies that $\Phi(x\otimes y)\neq 0$. Therefore (ii) implies (iii).

 Conversely, assume (iii) and let  $\{F_j\}_{j\in\Bbb{J}}$ be a POVM such that $ span_{j\in\Bbb{J}}\{F_j\} = B(H_B)$. 
Suppose to the contrary that (ii) is false. Then there exists $y\neq 0$ and $span_{j\in\Bbb{J}}\{\Phi(F_j)^*y\}\neq H_A$, and hence there exists $x \neq 0$ such that $x\perp span_{j\in\Bbb{J}}\{\Phi(F_j)^*y\}$. Thus for $\forall j\in\Bbb{J}$, we have $$0=\langle x,\Phi^*(F_j)y\rangle =tr((x\otimes y)(\Phi^*(F_j^*)))=\langle x\otimes y,\Phi^*(F_i)\rangle 
		=\langle \Phi(x\otimes y),F_j\rangle.$$
Since $ span_{j\in\Bbb{J}}\{F_j\} = B(H_B)$, we get that  $\Phi(x\otimes y)=0$, which leads to a contradiction to our assumption. Therefore (iii) implies (ii).
\end{proof}

In what follows, we say that a subset $\Omega$ of $\{1, ... , k\}$ is an ordered subset if $\Omega$ is listed in the increasing order.  As for an ordered bipartition of $\{1, ... , k\}$, we mean that both $\Omega$ and $\Omega^c$ are ordered subsets. For a $k$-tuple $\overrightarrow{A} = (A_1, ..., A_k)^t$ and an ordered index subset $\Omega = \{i_1, ... , i_\ell\}$ of $\{1, ... , k\}$, we write $\overrightarrow{A}_{\Omega} = (A_{i_1}, ..., A_{i_\ell})^t$.

\begin{theo} \label{main-thm-2} Let $\Phi: B(H_A)\to B(H_B)$ be a real quantum channel of Choi's rank-$r$  with Kraus operators $\overrightarrow{A} =  (A_1, ..., A_r)^{t}$. Then the following statements are equivalent

(i) $\Phi$ is phase retrievable.

(ii) For any ordered bipartition  $\Omega$ and $\Omega^c$ of  $\{1, ... , r\}$ and any $k\times (r-k)$ matrix  $\Lambda$,  we have either $\Lambda\notin\sigma_{\ell}(\overrightarrow{A}_{\Omega},\overrightarrow{A}_{\Omega^c})$ or $-\Lambda^*\notin\sigma_{\ell}(\overrightarrow{A}_{\Omega^c}, \overrightarrow{A}_{\Omega})$,
where $|\Omega| = k$.

Moreover, the implication $(i)\Rightarrow (ii)$ is also true for complex Hilbert spaces case.
\end{theo}

\begin{proof} By  Lemma \ref{lem-2.1}, it suffices to show that (ii) is equivalent to the condition that  $\Phi(x\otimes y) \neq 0$ whenever $x\otimes y \neq 0$. We will show next that this is actually true for both real and complex Hilbert spaces.

%
%
%
%
%
%
%

Assume that (ii) is false. Then there exists a bipartition $P=\{\Omega, \Omega^c\}$ and a $k \times (r-k)$ matrix $\Lambda$ such that $\Lambda\in\sigma_{\ell}(\overrightarrow{A}_{\Omega},\overrightarrow{A}_{\Omega^c})$ and $-\Lambda^*\in\sigma_{\ell}(\overrightarrow{A}_{\Omega^c}, \overrightarrow{A}_{\Omega})$. Without losing the generality, we can assume that $\Omega = \{1, ... , k\}$ and $\Omega^c = \{k+1, ... , r\}$.

By Lemma \ref{lem-spectrum}, there exist nonzero vectors $x, y\in H_{A}$ such that  $A_ix	=\sum_{j=r+1}^k\lambda_{ij}A_jx$ when $1\leq i\leq k$ and $A_jy= \sum_{i=1}^r-\bar{\lambda}_{ji}A_iy$ when $k+1\leq j\leq r$. Then we have
\begin{equation}
	\begin{aligned}
		\Phi(x\otimes y)&=\sum_{i=1}^{r}A_i(x\otimes y)A_i^*\\
		&=\sum_{i=1}^{k}A_i(x\otimes y)A_i^*+\sum_{i=k+1}^{r}A_i(x\otimes y)A_i^*\\
		&=\sum_{i=1}^{k}(\sum_{j=k+1}^{r}\lambda_{ij}A_jx_1)\otimes (A_ix_2)+\sum_{i=k+1}^{r}(A_ix)\otimes(\sum_{j=1}^{k}-\bar{\lambda}_{ji}A_jx_2)\\
		&=\sum_{i=1}^{k}\sum_{j=k+1}^{r}\lambda_{ij}(A_jx)\otimes (A_iy)-\sum_{i=k+1}^{r}\sum_{j=1}^{k}\lambda_{ji}(A_ix)\otimes (A_jy)\\
		&=\sum_{i=1}^{k}\sum_{j=k+1}^{r}\lambda_{ij}(A_jx)\otimes (A_iy)-\sum_{i=1}^{k}\sum_{j=k+1}^{r}\lambda_{ij}(A_jx)\otimes (A_iy)\\
		&=0.
	\end{aligned}
\end{equation}
Therefore we have proved that (ii) holds if $\Phi(x\otimes y) \neq 0$ whenever $x\otimes y \neq 0$.

Conversely, assume that there exist $x,y\neq 0$ such that $\Phi(x\otimes y)= 0$. First we show that $A_1x,\cdots,A_rx$ are linearly dependent. Indeed, if they are linearly independent, then there exist $z_j\in H_A$ such that $\langle z_j, A_ix\rangle = \delta_{ij}$, where $\delta_{ij}$ is the  Kronecker delta symbol. Note that $\sum_{i=1}^{r}A_{i}y\otimes A_ix = (\Phi(x\otimes y))^* = 0$. Thus for each $j$ we have
$$
0 =( \Phi(x\otimes y))^*z_{j} = (\sum_{i=1}^{r}A_{i}y\otimes A_ix)z_{j} =\sum_{i=1}^{r} \langle z_j, A_ix\rangle A_iy = A_{j}y,
$$
which is impossible since $\sum_{j=1}^{r}A_{j}^*A_{j} = I$. Therefore $A_1x,\cdots,A_rx$ are linearly dependent. Without losing the generality we can assume that $\{A_{k+1}x,\cdots,A_rx\}$ is a basis for $span \{A_1x,\cdots,A_rx\}$ and let $\Omega = \{1, ..., k\}$. Then there exist a $k\times (r-k)$ matrix $\Lambda$ such that 
$$
\left( \begin{matrix}A_1x\\ \vdots\\ A_kx\end{matrix} \right) = \Lambda \left( \begin{matrix}A_{k+1}x\\ \vdots\\ A_rx\end{matrix} \right).
$$
This implies by Lemma \ref{lem-spectrum}  that $\Lambda \in \sigma_{\ell}(\overrightarrow{A}_{\Omega},\overrightarrow{A}_{\Omega^c})$. Next we show that  $-\Lambda^*\in\sigma_{\ell}(\overrightarrow{A}_{\Omega^c}, \overrightarrow{A}_{\Omega})$.

Write $A_{j}x = \sum_{i=1}^{r-k}\lambda_{ji}A_{k+i}x$ for $j=1, ... ,k$. Then we get

\begin{align*}
0 &= \Phi(x\otimes y)=\sum_{i=1}^rA_i(x\otimes y)A_i^*\\
	&=\sum_{j=1}^kA_jx\otimes A_j y+\sum_{i=1}^{r-k}A_{k+i}x\otimes A_{k+i}y\\
	&=\sum_{j=1}^k(\sum_{i=1}^{r-k}\lambda_{ji}A_{k+i}x )\otimes A_j y+\sum_{i=1}^{r-k}A_{k+i}x\otimes A_{k+i}y\\
	&=\sum_{i=1}^{r-k}\sum_{j=1}^{k}\lambda_{ji}A_{k+i}x \otimes A_j y+\sum_{i=1}^{r-k}A_{k+i}x\otimes A_{k+i}y\\
	&=\sum_{i=1}^{r-k}\sum_{j=1}^{k}A_{k+i}x \otimes\bar{\lambda}_{ji} A_j y+\sum_{i=1}^{r-k}A_{k+i}x\otimes A_{k+i}y\\
	&=\sum_{i=1}^{r-k}A_{k+i}x \otimes\sum_{j=1}^{k}\bar{\lambda}_{ji} A_j y+\sum_{i=1}^{r-k}A_{k+i}x\otimes A_{k+i}y\\
	&=\sum_{i=1}^{r-k}A_{k+i}x \otimes(\sum_{j=1}^{k}\bar{\lambda}_{ji} A_j y+A_{k+i}y).\\
\end{align*}
Since $\{A_{k+1}x,\cdots,A_rx\}$ is linearly independent, we get from the above that $$\sum_{j=1}^{k}\bar{\lambda}_{ji} A_j y+A_{k+i}y =0$$ for every $i=1,... , r-k$, which implies by Lemma \ref{lem-spectrum} again that  $-\Lambda^*\in\sigma_{\ell}(\overrightarrow{A}_{\Omega^c}, \overrightarrow{A}_{\Omega})$. This completes the proof.
\end{proof}

As  stated in the proof above we actually have proved the following:

\begin{coro}  \label{coro-2.1} Let $\Phi: B(H_A)\to B(H_B)$ be a (real or complex) quantum channel of Choi's rank-$r$  with Kraus operators $\overrightarrow{A} =  (A_1, ..., A_r)^t$. Then the following are equivalent:

(i) $\Phi(x\otimes y) \neq 0$ whenever $x\otimes y \neq 0$.

(ii) For any ordered bipartition any bipartition  $\Omega$ and $\Omega^c$ of  $\{1, ... , r\}$ and any matrix  $\Lambda\in M_{k\times (r-k)}(\Bbb{R})$,  we have either $\Lambda\notin\sigma_r(\overrightarrow{B}_{\Omega},\overrightarrow{B}_{\Omega^c})$ or $-\Lambda^*\notin\sigma_r(\overrightarrow{B}_{\Omega^c}, \overrightarrow{B}_{\Omega})$,
where $ k = |\Omega|=r$.
\end{coro}

\begin{rem} While Theorem \ref{main-thm-1} tells us that  Theorem \ref{main-thm-2}  is also true for complex Hilbert spaces when $Cr(\Phi) =2$,
 the next  example shows that when $Cr(\Phi) > 2$, condition (ii) in  Theorem \ref{main-thm-2} is no longer sufficient for the phase-retrievability of $\Phi$ in the complex case.
\end{rem}

\begin{exam}  By Proposition \ref{prop-1.1} there exist $n\geq 2$ and  a frame $\{f_j\otimes f_j\}_{j=1}^{N}$ for $\C^n$  such that it has the complement property,  not phase retrievable and $\{f_j\otimes f_j\}_{j=1}^{N}$ linearly independent. Such an example can be easily constructed (For example, let $f_1 = e_1, f_2 = e_2$ and $f_3 = e_1+ e_2$, where $\{e_1, e_2\}$ is the standard basis for $\Bbb{C}^2$). The complement property of the frame implies that $N\geq 2n-1 \geq 3.$ 
 Let $A_j = f_j\otimes f_j$ and $S = \sum_{j=1}^{N}A_{j}A_{j}^*$. The operator $S$ is positive and invertible. Define $\Phi(T) = \sum_{j=1}^{N}A_jTA_j^*$ and $\tilde{\Phi}(T) = \sum_{j=1}^{N}B_jTB_j^*$, where $B_j = S^{-1/2}A_j$. Then $\tilde{\Phi}$ is a quantum channel. 
 
 If $\tilde{\Phi}(x\otimes y) = 0$, then clearly $\Phi(x\otimes y) = 0$. Thus we get 
$$
\sum_{j=1}^{N}  \langle x, f_j\rangle \langle f_j, y \rangle  f_j\otimes f_j = 0,
$$
which implies that $ \langle x, f_j\rangle \langle f_j, y \rangle  = 0$ for every $j$. Therefore we have $y \perp span\{ (f_j\otimes f_j)x: 1\leq j\leq N\}$. Since $\{f_j\}_{j=1}^{N}$ has the complement property, we get by Proposition \ref{prop-1.1} (i) that either $x=0$ or $y = 0$ and hence $x\otimes y =0$.
Therefore,  by Corollary \ref{coro-2.1}, we have that $\tilde{\Phi}$ satisfies the condition (ii) in Theorem \ref{main-thm-2}
. However, since $\{f_j\}_{j=1}^{N}$ is not phase retrievable, there exist $x, y\in H$ such that $\rho_x \neq \rho_y$ but $|\langle x, f_j\rangle | = |\langle y,  f_j\rangle |$ for all $j$. This implies 
$$
\Phi(x\otimes x) = \sum_{j=1}^{N}  |\langle x, f_j\rangle|^2   f_j\otimes f_j  = \sum_{j=1}^{N}  |\langle y, f_j\rangle |^2  f_j\otimes f_j = \Phi(y\otimes y),
$$
and consequently $\tilde{\Phi}(x\otimes x) = \tilde{\Phi}(y\otimes y)$.
Therefore, by Proposition \ref{prop-1.3}, we get that $\tilde{\Phi}$ is not phase retrievable.
\end{exam}

\subsubsection{The complex Hilbert space case}

Similar to Lemma \ref{lem-2.1}, for complex quantum channels, we have

\begin{lem} \label{lem-2.3} Let $\Phi: B(H_A)\to B(H_B)$ be a quantum channel over complex Hilbert spaces $H_A$ and $H_B$. Then the following are equivalent:
	
	(i) $\Phi$ is phase retrievable.
	
	(ii)  $\Phi(x \otimes y+ y\otimes x)\neq 0$ whenever $x \otimes y+ y\otimes x \neq 0$.
\end{lem}
\begin{proof}
$(i) \Rightarrow (ii)$: Suppose $\Phi$ does phase retrieval. Then there exists a POVM $\{F_j\}_{j\in\Bbb{J}}$ on $H_B$ such that  $\{\Phi^*(F_j)\}_{j\in\Bbb{J}}$ does phase retrieval for $H_A$. That is: if $$\langle \Phi^*(F_j),x\otimes x\rangle =\langle \Phi^*(F_j),y\otimes y\rangle $$ for $\forall j\in\Bbb{J},$ then $x\otimes x = y\otimes y$.
Now assume that  $x \otimes y+ y\otimes x \neq 0$. If $y = cx$  for some $c \neq 0$, then we have $c + \bar{c} \neq 0$. This implies that
$$
\Phi(x \otimes y+ y\otimes x) = (c + \bar{c}) \Phi(x\otimes x) \neq 0.
$$
If $ x$ and $ y$ are linear independent, then $x+y$ and $x-y$  are also linear independent. Since $\{\Phi^*(F_j)\}_{j\in\Bbb{J}}$ is phase retrievable for $H_A$, we get that 
 there exists $j_0\in\Bbb{J}$ such that $$\langle \Phi^*(F_{j_0}),(x+y)\otimes (x+y)\rangle \neq \langle \Phi^*(F_{j_0}),(x-y)\otimes (x-y)\rangle,$$ which implies that $$\langle F_{j_0},\Phi(x\otimes y+y\otimes x) \rangle = \langle \Phi^*(F_{j_0}),(x\otimes y+y\otimes x)\rangle \neq 0.$$
Therefore $\Phi(x\otimes y+y\otimes x)\neq 0$.\\
\\
$(ii) \Rightarrow (i)$: Let $\{F_j\}_{j\in\Bbb{J}}$ be a POVM such that $span\{F_j: j\in \Bbb{J}\} = B(H_{B})$. If $\Phi$ is not phase retrievable, then there exist two vectors $x, y\in H_A$ linearly independent such that
	$\langle\Phi^*(F_j),x\otimes x\rangle =\langle \Phi^*(F_j),y\otimes y\rangle$ for every $j\in \Bbb{J}$.

Let $u = x+ y$ and $v = x-y$. Then $u\otimes v + v\otimes u = 2(x\otimes x - y\otimes y) \neq 0$. However,  for every $j\in\Bbb{J}$, we have
$$
\langle F_j, \Phi(u\otimes v + v\otimes u) \rangle  =\langle \Phi^*(F_j), u\otimes v + v\otimes u\rangle  = 2\langle \Phi^*(F_j), x\otimes x - y\otimes y \rangle =0.
$$
This implies that  $\Phi(u\otimes v + v\otimes u) = 0$, which contradicts with our assumption (ii). Therefore $\Phi$ is phase retrievable. 
\end{proof}



%
%
%
%

\begin{defi} Two  vectors $u= (u_1, ... , u_k), v = (v_1, ... v_k)\in \oplus_{j=1}^{k}H$ are {\it skew-commutative} if $\sum_{j=1}^{k}(u_j\otimes v_j + v_{j}\otimes u_{j}) = 0$, i.e., $A = \sum_{j=1}^{k}u_j\otimes v_{j}$ is skew-Hermitian. We say that two subsets $U$ and $V$ of $ \oplus_{j=1}^{k}H$ are {\it completely non-skew commutative } if any two nonzero vectors $u\in U$ and $v\in V$ are not skew-commutative.
\end{defi}

In what follows we adopt the following notations: For $\overrightarrow{A} = (A_1, ... , A_k)^{t}$ with $A_{j}\in B(H, K)$,  we write $\overrightarrow{A}x = (A_1x, ... , A_kx)^t $ for $x\in H$. Then
$Ker \overrightarrow{A} = \{x\in H: \overrightarrow{A}x = 0\}$ and 
$\overrightarrow{A}(M) = \{\overrightarrow{A}x: x\in M\}$, where $M$ is a subset of $H$.


\begin{prop}  Let $\Phi: B(H_A)\to B(H_B)$ be a complex quantum channel of Choi's rank-$r$  with Kraus operators $\overrightarrow{A} =  (A_1, ..., A_r)^t$. If $\Phi$ is not phase retrievable, then $ \overrightarrow{A_{\Omega_c}}(M)$ and $ (\Lambda^*\overrightarrow{A_{\Omega}}+\overrightarrow{A_{\Omega_c}})(M^c)$ are completely non-skew commutative for any ordered bipartition  $\Omega$ and $\Omega^c$ of  $\{1, ... , r\}$ and any matrix $\Lambda\in M_{k\times (r-k)}(\Bbb{C})$, where $ k = |\Omega|$, $M = Ker(\overrightarrow{A_{\Omega}}-\Lambda\overrightarrow{A_{\Omega_c}}))$ and $M^c = \{x\in H_{A}: x\notin M\}$.
\end{prop}

\begin{proof} 
Suppose  to the contrary that there is an ordered  bipartition $\{\Omega, \Omega^c\}$, say $\Omega = \{1, .. , k\}$ and a matrix $\Lambda\in M_{k\times (r-k)}(\Bbb{C})$ such that 
 $ \overrightarrow{A_{\Omega_c}}(M)$ and $ (\Lambda^*\overrightarrow{A_{\Omega}}+\overrightarrow{A_{\Omega_c}})(M^c)$ are not completely non-skew commutative. Then there exist nonzero vectors $u\in \overrightarrow{A_{\Omega_c}}(Ker(\overrightarrow{A_{\Omega}}-\Lambda\overrightarrow{A_{\Omega_c}}))$ and $v\in (\Lambda^*\overrightarrow{A_{\Omega}}+\overrightarrow{A_{\Omega_c}})(M^c)$ that are skew-commutative. Therefore
 there exists  a nonzero vector $x$ such that $u=\overrightarrow{A_{\Omega_c}}x$ and $(\overrightarrow{A_{\Omega}}-\Lambda\overrightarrow{A_{\Omega_c}})x=0$, i.e.,  $u_i= A_{k+i}x$ for $i =1, ... r-k$ and $A_{j}x= \sum_{i=1}^{r-k} \lambda_{ji}A_{k+i}x$ for  every $1\leq j\leq k$. 
Similarly,  there exists a nonzero vector  $y\in M^c$ such that $v=(\Lambda^*\overrightarrow{A_{\Omega}}+\overrightarrow{A_{\Omega_c}})y$, i.e.,  $v_i=\sum_{j=1}^k\bar{\lambda}_{ji}A_jy+A_{k+i}y$ for $i=1, ..., r-k$. Since $x\in M$ and $y\in M^c$, we get that $x, y$ are linearly independent and hence $x\otimes y + y\otimes x \neq 0$.  Moreover
$
\sum_{j}^{k} v_j\otimes u_j = \Phi(x\otimes y).
$
Thus $\Phi(y\otimes x) = (\Phi(x\otimes y))^* = \sum_{j=1}^{k}u_j\otimes v_j$. Therefore we get 
$$
\Phi(x\otimes y + y\otimes x) = \sum_{j=1}^{k}v_j\otimes u_j + u_j\otimes v_j = 0
$$ and $x\otimes y + y\otimes x \neq 0$. This implies by Lemma 2.7 that $\Phi$ is not phase retrievable. This completes the proof.
\end{proof}

Finally we present a  relatively simple necessary condition for quantum channels to be phase retrievable. This condition  is slightly different from the ones we have discussed above but similar to the one in the case when $Cr(\Phi) = 2$. 

Given a list of operators $A_1, ... , A_r \in B(H_A, H_B)$, we will use $\sigma_{A_j,\ell}(A_1, ..., A_{j-1}, A_{j+1}, ... , A_r) $ to denote the set of all $\lambda = (\lambda_1, ..., \lambda_{j-1}, \lambda_{j+1}, ..., \lambda_{r})^t \in \C^{r-1}$ such that $$(A_1-\lambda_1A_j, ... , A_{j-1}-\lambda_{j-1}A_j, A_{j+1}-\lambda_{j+1}A_{j}, ... , A_r-\lambda_rA_j)^t$$ is not left invertible.

\begin{prop} Let $\Phi: B(H_A)\rightarrow B(H_B)$ be a quantum channel with Kraus operators $A_1 , ... , A_r$. If $\Phi$ is phase retrievable,  then 
$$
1 + \langle \lambda , \mu\rangle \neq 0.
$$
for every $j\in\{1,2,\cdots,r\}$ and any $\lambda,\mu\in \sigma_{A_j,\ell}(A_1, ..., A_{j-1}, A_{j+1}, ... , A_r)$.
\end{prop}
 \begin{proof} 
 Let $\{F_j=f_j\otimes f_j\}_{j\in\Bbb{J}}$ be  a POVM in $B(H_B)$ such that $\{\Phi^*(F_j)\}_{j\in\Bbb{J}}$ does phase retrieval for $H_A$. Assume to the contrary that there exists a $j\in\Bbb{J}$, say $j=1$ and 
  $\lambda,\mu\in \sigma_{A_1,\ell}(A_2,  ... , A_r)$ such that 
  $$
  1+ \langle \lambda , \mu\rangle = 0.
  $$
 
Since $\lambda=(\lambda_2,\cdots,\lambda_r)\in\sigma_{A_1,\ell}$, there exist two nonzero  vectors $x, y\in H_A$ such that $A_ix=\lambda_iA_1x$ and  $A_iy=\mu_iA_1y$ for $i=2, ... , r$. Thus for every $F_j$ we get
\begin{equation*}
	\begin{aligned}
		\langle \Phi^*(F_j)x, y \rangle &=\sum_{i=1}^r\langle A_i^*F_jA_ix,y\rangle=\sum_{i=1}^r\langle F_jA_ix,A_iy\rangle\\
		&=\langle F_jA_1x,A_1y\rangle+\sum_{i=2}^r\langle F_j(\lambda_iA_1x),F_j(\mu_iA_1y)\rangle \\
		&=\langle F_jA_1x,A_1y\rangle +\sum_{i=2}^r\lambda_i\bar{\mu}_i\langle F_jA_1x,A_1y\rangle \\
		&=(1+\lambda_2\bar{\mu}_2+\cdots+\lambda_r\bar{\mu}_r)\langle F_jA_1x,A_1y\rangle = 0\\
	\end{aligned}
\end{equation*}
This implies that $span_{j\in\Bbb{J}}\{\Phi^*(F_j)x\}\neq H_B$, and hence $\Phi$ can not  be phase retrievable. 
 \end{proof}
 
 \begin{exam} Let $A_1=\frac{1}{\sqrt{3}}\begin{pmatrix}
		1&0\\0&1\end{pmatrix}$, $A_2=\frac{1}{\sqrt{3}}\begin{pmatrix}
		0&1\\1&0\end{pmatrix}$ and $A_3=\frac{1}{\sqrt{6}}\begin{pmatrix}
		1&-1\\-1&1\end{pmatrix}$ and $\Phi$ be the quantum channel represented by Kraus operators $\{A_1,A_2,A_3\}$. Then $\sigma_{A_1,\ell}(A_2, A_3)=\{(-1,\sqrt{2})^t,(1,0)^t\}$, and hence $\Phi$ is not phase retrievable.
\end{exam}

\section{Quantum Channels with Minimal Number of Observables}

In this section we consider the problem of constructing phase retrievable quantum channels that only require minimal number of observables. For an $n$-dimensional Hilbert space $H$, we will use $d_n$ to denote the smallest integer $N$ such that there is a phase retrievable frame $\{x_{j}\}_{j=1}^{N}$ for $H$. By Proposition \ref{prop-1.1}, $d_n = 2n-1$ if $H$ is a real Hilbert space, and $d_n \leq  4n-4$ if $H$ is a complex Hilbert space. In the special case when $n = 2^k+1$ and $H$ is a complex Hilbert space, we have that $d_n = 4n-4$.

\begin{lem}  \label{lem-3.1} Let $H$ be an $n$-dimensional Hilbert space. If  $\{f_j\}_{j=1}^{d_n}$ is a phase retrievable frame for $H$, then $\{f_j\otimes f_j\}_{j=1}^{d_n}$ is linearly independent.
\end{lem} 
\begin{proof} Assume otherwise that $\mathcal{F}: = \{f_j\otimes f_j\}_{j=1}^{d_n}$ is linearly dependent. Then we can assume that $f_{1}\otimes f_{1} \in span \{f_j\otimes f_j: 2\leq j\leq d_n\}$. Let $\mathcal{G} =  \{f_j\otimes f_j\}_{j=2}^{d_n}$. Then $\ker L_{\mathcal{F}} = \ker L_{\mathcal{G}}$. Thus, by Proposition \ref{prop-1.1} (v) we get that $\mathcal{G}$ is also a phase retrievable frame for $H$. This contradicts with the minimality of $d_n$. Thus $\{f_j\otimes f_j\}_{j=1}^{d_n}$ is linearly independent.
\end{proof}

 In what follows, we say that a phase retrievable quantum channel $\Phi: B(H_A) \to B(H_B)$ admits the minimal number of  rank-one observables if there exists $d_n$-number of rank-one observables $\{f_j\otimes f_j\}_{j=1}^{d_n}$ such that $\{\Phi^*(f_j\otimes f_j)\}_{j\in\Bbb{J}}$ does phase retrieval for $H_A$, where $n = \dim H_A$. While we are not able to characterize all the quantum channels that admit the minimal number of rank-one observables, we present  some cases  (Corollary 3.4 and Corollary 3.7) where  phase retrievable quantum channels of rank $r$ with the minimal number of rank-one observable property can be easily constructed. 

We first examine the case when $Cr(\Phi) = 2$.  Let $\Phi: B(H_{A}) \to B(H_{B})$ be a quantum channel with Kraus operators $A_1, A_2 \in B(H_{A}, H_{B})$. If $\Phi$ is phase retrievable, then by Theorem \ref{main-thm-1} we get that either $A_1$ or $A_{2}$ is injective. 

\begin{prop} \label{prop-3.1} If $A_1$ is injective and $A_2$ is a rank-one operator, then $\Phi$ is phase retrievable that admit the minimal number of observables.
\end{prop} 

\begin{proof} Write $A^*_2 = u\otimes v$, where $u\in H_A$ and $v\in H_B$ are nonzero vectors. Since $A_1^*$ is surjective, there exists $f_1\in H_B$ such that $u = A_1^*f_1$. Then we can pick $d_n-1$ vectors $u_2, ... , u_{d_n}\in H_A$ such that  $\{u, u_2, ... , u_{d_n}\}$ is a phase retrievable frame for $H_A$. Write $u_j = A^*_{1}f_j$ for some $f_j\in H_B$ and $j =2, ..., d_n$. Then 
$$
\Phi^*(F_1) = A_{1}^*f_1 \otimes A_{1}^{*}f_1 +  A_{2}^*f_1 \otimes A_{2}^{*}f_1 = (1 + | \langle v, f_1\rangle |^2) u\otimes u
$$ and
$$
\Phi^*(F_j) = A_{1}^*f_j \otimes A_{1}^{*}f_j +  A_{2}^*f_j \otimes A_{2}^{*}f_j  = u_j\otimes u_j +  | \langle v, f_1\rangle |^2 u\otimes u
$$
for $j =2, ..., d_n$.

Now suppose that $\langle x, \Phi^*(F_j)x\rangle  = \langle y, \Phi^*(F_j)y\rangle $ for all $j$. Note that 
$$
(1 + | \langle v, f_1\rangle |^2)| \langle x, u\rangle|^2  = (1 + | \langle v, f_1\rangle |^2) | \langle y, u\rangle|^2 
$$
and
$$
|\langle x, u_j\rangle |^2 +  | \langle v, f_1\rangle |^2| \langle x, u\rangle|^2  = |\langle y, u_j\rangle |^2 + | \langle v, f_1\rangle |^2 | \langle y, u\rangle|^2 
$$
for $j =2, ... , d_n$. The above two identities combined  implies  that $\langle x, u\rangle|^2  = | \langle y, u\rangle|^2$ and $ |\langle x, u_j\rangle |^2 = |\langle y, u_j\rangle |^2$ for $j\geq 2$. Since $\{u, u_2, ... , u_{d_n}\}$ is phase retrievable, we immediately get that $x\otimes x = y\otimes y$.
\end{proof}


\begin{rem} The above example  does not hold when $rank(A_2) \geq 2$,  or when $r= Cr(\Phi) > 2$ with Kraus operators $\{A_1, ... , A_r\}$ such that $A_1$ is injective and $A_2, ... , A_{r}$ are all rank-one operators. Here are two simple examples:  Let $H_A = H_B = \C^2$.

(i)  $A_1  = {1\over \sqrt{2}} I$ and $ A_2 =  {1\over \sqrt{2}} (e_1\otimes e_2 + e_2\otimes e_1)$. 

(ii) $A_1  = {1\over \sqrt{2}} I, A_2 =  {1\over \sqrt{2}} e_1\otimes e_2$ and $A_3 =  {1\over \sqrt{2}} e_2\otimes e_1$.

Since in either case we have $\Phi(e_1\otimes e_1) = \Phi(e_2\otimes e_2) = I$ and hence $\Phi$ is not phase retrievable. 
\end{rem}

For the case when  $Cr(\Phi) >2$,  we next present a sufficient condition on the rank-one Kraus operators  under which  $\Phi$  is a phase retrievable quantum channel that admits the minimal number of observables. As a consequence,  this is true when one of $\{A_1, ..., A_r\}$ is positive invertible and the rest are positive rank-one operators.

 
 \begin{prop} \label{prop-3.2}  Let $\Phi: B(H_A) \to B(H_B)$ be a quantum channel of rank-$r$ with Kraus operators $\{A_1, ... , A_r\}$ 
 such that $A^*_r$ is surjective and $A_j^* = u_j\otimes v_j$ with $u_j= A_r^*f_j$ for  some $f_j\in H_B$, $j=1, ..., r-1$. If the $(r-1)\times (r-1)$-matrix $I  + [|\langle v_i, f_j\rangle|^2]$ is invertible, then $\Phi$ is phase retrievable. Moreover, if in addition assume that $\{u_j\otimes u_j\}_{j=1}^{r-1}$ is linearly independent, then there exist $d_n$ number of rank-one observables $F_j = f_j\otimes f_j\in B(H_B)$ such that $\{\Phi^*(F_j)\}_{j=1}^{d_n}$ is phase retrievable for $H_{A}$ where $n = \dim H_{A}$.
\end{prop} 

\begin{proof}  Let $M$ be the matrix $I  + [|\langle v_i, f_j\rangle|^2]$ in $M_{(r-1)\times(r-1)}(\Bbb{C})$ and assume that $M$ is invertible. Extend $\{u_j\}_{j=1}^{r-1}$ to a phase retrievable frame $\{u_j\}_{j=1}^{N}$ for $H_A$. For $j\geq r$, write $u_j = A^*_r f_j$ for some $f_j\in H_B$, and then let $F_j = f_j\otimes f_j$ for $1\leq j\leq d_n$. Note that 
 \begin{align*}
 \langle x, \Phi^*(F_j)x\rangle  &=\sum_{i=1}^{r}| \langle x, A^*_if_j\rangle |^2\\
 & = \sum_{i=1}^{r-1}|\langle v_i, f_{j}\rangle |^2 |\cdot  |\langle x, u_i \rangle |^2 + |\langle x, u_j \rangle |^2.\\
\end{align*}
Thus we get
$$
\left[\begin{array}{c} \langle x, \Phi^*(F_1)x\rangle\\
 \langle x, \Phi^*(F_2)x\rangle\\
 \vdots \\
 \langle x, \Phi^*(F_{r-1})x\rangle\\
\end{array}\right] = M \left[\begin{array}{c} |\langle x, u_1\rangle |^2\\
 |\langle x, u_2\rangle |^2\\
 \vdots \\
 |\langle x, u_{r-1}\rangle |^2.
 \end{array}\right] 
$$
Now suppose that  $\langle x, \Phi^*(F_j)x\rangle  = \langle y, \Phi^*(F_j)y\rangle$ for all $1\leq j\leq N$. Then the above argument implies that 
 $$
 M \left[\begin{array}{c} |\langle x, u_1\rangle |^2\\
 |\langle x, u_2\rangle |^2\\
 \vdots \\
 |\langle x, u_{r-1}\rangle |^2.
 \end{array}\right] = M \left[\begin{array}{c} |\langle y, u_1\rangle |^2\\
 |\langle y, u_2\rangle |^2\\
 \vdots \\
 |\langle y, u_{r-1}\rangle |^2.
 \end{array}\right] 
$$
and hence  $|\langle x, u_j\rangle |^2 =  |\langle y, u_j\rangle |^2$ for $1\leq j\leq r-1$ since $M$ is invertible. In particular, this implies that  $$\sum_{i=1}^{r-1}|\langle v_i, f_{j}\rangle |^2 |\cdot  |\langle x, u_i \rangle |^2 =  \sum_{i=1}^{r-1}|\langle v_i, f_{j}\rangle |^2 |\cdot  |\langle y, u_i \rangle |^2 $$  holds for all $j$.
Therefore, if  $j\geq r$,  then from $$\sum_{i=1}^{r-1}|\langle v_i, f_{j}\rangle |^2 |\cdot  |\langle x, u_i \rangle |^2 + |\langle x, u_j \rangle |^2 =  \sum_{i=1}^{r-1}|\langle v_i, f_{j}\rangle |^2 |\cdot  |\langle y, u_i \rangle |^2 + |\langle y, u_j \rangle |^2$$, we get that $  |\langle x, u_j \rangle |^2 = |\langle y, u_j \rangle |^2$. Consequently we have $  |\langle x, u_j \rangle |^2 = |\langle y, u_j \rangle |^2$ for every $j =1, ... , N$ and hence $x\otimes x = y\otimes y$. Thus $\Phi$ is phase retrievable. 

If in addition assume that $\{u_j\otimes u_j\}_{j=1}^{r-1}$ is linearly independent,  then we we can extend  $\{u_j\}_{j=1}^{r-1}$ to a phase retrievable frame $\{u_j\}_{j=1}^{d_n}$ for $H_A$ (i.e., $N = d_n)$. The above proof shows that $\{\Phi^*(F_j)\}_{j=1}^{d_n}$ is phase retrievable for $H_{A}$.
 \end{proof}

\begin{coro} \label{cor-3.1} Let $\Phi: B(H) \to B(H)$ be a quantum channel  of rank-$r$ with positive  Kraus operators $\{A_1, ... , A_r\}$ such that $A_r$ is invertible and $A_1, ... , A_{r-1}$ are rank-one operators.   Then $\Phi$ is phase retrievable. Moreover, there exists $d_n$ number of rank-one observables $F_j = f_j\otimes f_j\in B(H_B)$ such that $\{\Phi^*(F_j)\}_{j=1}^{d_n}$ is phase retrievable for $H_{A}$, where $n = \dim H_{A}$.
\end{coro}
\begin{proof} Write $A_j = u_j\otimes u_j$ and $u_j = A_r f_j$ for $j =1, ... , r-1$. Since $Cr(\Phi) = r$, we get that $\{u_j\otimes u_j\}_{j=1}^{r-1}$ is linearly independent. Moreover,  the matrix $[ |\langle u_i, f_j\rangle|^2] =  [|\langle A_{r}^{1/2}f_i,  A_{r}^{1/2}f_j\rangle|^2]$ is the Gramian matrix of the vectors $( A_{r}^{1/2}f_1\otimes  A_{r}^{1/2}f_1, ...  , A_{r}^{1/2}f_{r-1}\otimes A_{r}^{1/2}f_{r-1})$ in $B(H)$, which is positive semidefinite. Thus $I  + [|\langle v_i, f_j\rangle|^2]$ is invertible, and so the consequence follows from Proposition \ref{prop-3.2}

\end{proof}

The following is a negative result on phase retrievable quantum channels. 

\begin{prop} \label{prop-3.3} Let $\Phi: B(H_A) \to B(H_B)$ be a quantum channel of Choi rank-$r$ with Kraus operators $A_1, ... , A_r$. If $\sum_{i=1}^{r}range(A_i^*)$ is a direct sum and $r > 1$, then $\Phi$ is not phase retrievable. In particular,  $\Phi: B(H) \to B(H)$ is not phase retrievable if $\Phi(T) = \sum_{i=1}^{r}P_{i}TP_{i}$ with mutually orthogonal projections $P_1, ... , P_r$ and $r\geq 2$.

\end{prop} 
\begin{proof} Write $A_{i}^* = \sum_{j=1}^{k_i}g_{ij}\otimes f_{ij}$, where $\{g_{ij}\}_{j=1}^{k_i}$ is a basis  for $M_i : = range(A_i^*)$. Then $\{g_{ij}: 1\leq i\leq r, 1\leq j\leq k_i\}$ is a basis for $H_A$. Let $\{h_{ij}: 1\leq i\leq r, 1\leq j\leq k_i\}$ be its dual basis, and set $x = h_{11} + h_{22}$ and $y = h_{11} - h_{22}$. Then a direct calculation shows that
$$
\Phi(x\otimes x) = \sum_{i=1}^{r}\sum_{j, j' = 1}^{k_{i}} \langle x, g_{ij}\rangle  \langle g_{ij'}, x \rangle  f_{ij}\otimes f_{ij'} = f_{11}\otimes f_{11} + f_{21}\otimes f_{21}
$$
and
$$
\Phi(y\otimes y) = \sum_{i=1}^{r}\sum_{j, j' = 1}^{k_{i}} \langle y, g_{ij}\rangle  \langle g_{ij'}, y \rangle  f_{ij}\otimes f_{ij'} = f_{11}\otimes f_{11} + f_{21}\otimes f_{21}.
$$
Thus $\Phi$ is not phase retrievable.
\end{proof}

Next we consider the following problem: Given a list of linearly independent observables $\{F_j\}_{j=1}^{N}$ in $B(H_B)$. Under what condition does there exist a quantum channel $\Phi$ such that $\{\Phi^*(F_j)\}_{j=1}^N$ does phase retrieval for $B(H_A)$? If it does exist, then what can we say about the Choi's rank of such channels? Next result partially answered these questions for the rank-one observable case.

\begin{prop} \label{prop-3.4} Assume that $\dim H_A = \dim H_B$. Let $\{F_j\}_{j=1}^{N} = \{f_j\otimes f_j\}_{j=1}^{N}$ be linearly independent rank-one operators such that $\{f_j\}_{j=1}^{N}$ is a phase retrievable frame for $H_B$ and $I\notin span\{F_j: j\in \Lambda\}$ for any proper subset $\Lambda$ of $\{1, ... , N\}$. Then for any integer $r\in \{1, ... , N\}$, there is a quantum channel $\Phi$  of Choi's rank-$r$ such that $\{\Phi^*(F_j)\}_{j=1}^{N}$ does phase retrieval for $B(H_A)$.

\end{prop}

\begin{proof}  Let $U: H_A \to M$ be a unitary operator,  $g_j = U^*f_j$ and  $A_j = f_j\otimes g_j$.  Then $\{g_j\otimes g_j\}_{j=1}^{N}$ does phase retrieval for $H_A$.

(i) For $r = 1$, let $\Phi(T) = UTU^*$. Then $Cr(\Phi) = 1$ and $\{\Phi^*(F_j)\}_{j=1}^{N} =  \{ g_j\otimes g_j\}_{j=1}^{N}$ does phase retrieval for $H_A$.

(ii) For $2\leq r \leq N$, let $\Phi(T) =  \sum_{j=1}^{r-1}A_jTA_j^* + UTU^*$. We first claim that $Cr(\Phi) = r$. Indeed if $cU + \sum_{j=1}^{r-1}c_jA_j = 0$, then we get
$$
(cI + \sum_{j=1}^{r-1}(f_j\otimes f_j)) U = 0
$$
which implies $c = c_1 = ... c_{r-1} = 0$ since by assumption  $\{I, f_1\otimes f_1, ... , f_{r-1}\otimes f_{r-1}\}$ is linearly independent. Thus $Cr(\Phi) = r$. Next we claim that $\{\Phi^*(F_j)\}$ dose phase retrieval for $H_A$. Indeed, as it is shown in the proof of Proposition \ref{prop-3.2}, the $(r-1)\times (r-1) $ matrix $M =  I  + [|\langle f_i, f_j\rangle|^2]$ is always invertible. Thus $\langle x\otimes x, \Phi^*(F_j)\rangle = \langle y\otimes y, \Phi^*(F_j)\rangle$ for all $j$ imply that $x\otimes x = y\otimes y$. Hence  $\{\Phi^*(F_j)\}_{j=1}^{N}$ does phase retrieval for $B(H_A)$.
\end{proof}

\begin{coro} \label{coro-3.2} Let $H$ be an $n$-dimensional real Hilbert space and let $\{f_j\}_{j=1}^{d_n}$ be a phase retrievable Parseval frame for $H$. Then for every $1\leq r\leq d_n$, there exists of a quantum channel $\Phi$ of rank-$r$ such that $\{\Phi(f_j\otimes f_j)\}_{j=1}^{d_n}$ does phase retrieval for $H$. 
\end{coro}

\begin{proof} Since $\{f_j\}_{j=1}^{d_n}$ is a phase retrievable frame,  we get  from Lemma \ref{lem-3.1} that $\{f_j\otimes f_j\}_{j=1}^{d_n}$ is linearly independent.  By Proposition \ref{prop-3.4}, we only need to show  that $I\notin span\{F_j: j\in \Lambda\}$ for any proper subset $\Lambda$ of $\{1, ... , 2n-1\}$. If there is a proper subset $\Lambda$ such that $I\in span\{F_j: j\in \Lambda\}$. This implies that 
$$
\sum_{j\in\Lambda} f_{j}\otimes f_j = I = \sum_{j=1}^{d_n}f_{j}\otimes f_j, \ \  i.e., \ \  \sum_{j\in\Lambda} (1-c_j) f_{j}\otimes f_j + \sum_{j\in\Lambda^c} f_{j}\otimes f_j = 0
$$
which is impossible since $\{f_{j}\otimes f_j\}_{j\in \Lambda}$ is linearly independent and $\Lambda^c$ is nonempty. Therefore $I\notin span\{F_j: j\in \Lambda\}$ for any proper subset $\Lambda$ of $\{1, ... , 2n-1\}$.
\end{proof} 

\begin{rem} In the real Hilbert space case, every generic frame of length $N \geq  2n-1$ is phase retrievable. Therefore every generic Parseval frame of length $2n-1$ satisfies the condition of Corollary \ref{coro-3.2}.
\end{rem} 


\begin{thebibliography}{50}



\bibitem{14Akrami} F.  Akrami, P. G. Casazza, M. A. Hasankhani Fard, A. Rahimi, A note on phase (norm) retrievable Real Hilbert space (fusion) frames, arXiv preprint arXiv:2107.09606 (2021).

\bibitem{12Hasan} F.  Ali, and S.   Moazeni, Signal reconstruction without phase by norm retrievable frames, {\it Linear and Multilinear Algebra,}  69.8 (2021): 1484-1499.

\bibitem{Ba5} B. Balan, B.G.Bodmann, P.G. Casazza and D. Edidin, Painless Reconstruction from Magnitudes of Frame Vectors,  {\it J. Fourier Anal. Appl.,}  15 (2009), 488-501.

\bibitem{15Balan} R. Balan, Radu, P.  Casazza, and D.  Edidin, On signal reconstruction without phase,  {\it Applied and Computational Harmonic Analysis,}  20.3 (2006): 345-356.

\bibitem{16Balan} R. Balan, Stability of phase retrievable frames,  Wavelets and Sparsity XV. Vol. 8858. SPIE, 2013.

\bibitem{17Balan}  R. Balan, Frames and Phaseless Reconstruction,  AMS Short Course at the Joint Mathematics Meetings, San Antonio, January 2015. Proceedings of Symposia in Applied Mathematics 73 175-199 (2016).

\bibitem{Beneddeto} J.  Benedetto and A.  Kebo, The role of frame force in quantum detection, {\it Journal of Fourier Analysis and Applications,}  14(2008): 443--474.



\bibitem{10Bodmann} B. G Bodmann and J. Haas., A short history of frames and quantum designs, {\it Topological Phases of Matter and Quantum Computation,}  Contemporary Mathematics, 747 (2020): 215-226.

\bibitem{BCCHT2} S. Botelho-Andrade,  P. G. Casazza,  D. Cheng,  J. Haas, T. T. Tran, The Quantum Detection Problem: A Survey. In: Quantum Theory and Symmetries with Lie Theory and Its Applications in Physics Volume 2.  Springer Proceedings in Mathematics \& Statistics, vol 255. Springer, Singapore. (2018)

\bibitem{BCCHT1} S. Botelho-Andrade,  P. G. Casazza,  D.Cheng,  J. Haas,  T.T.Tran, The Solution to the Frame Quantum Detection Problem, {\it J. Fourier. Anal. Appl.,} 25(2019), 2268–2323.

\bibitem{Choi1} M. D. Choi, Completely positive linear maps on complex matrices,  {\it Linear Algebr. Appl.,} 10 (1975): 285--290.

\bibitem{CEHV-ACHA} A. Conca, D. Edidin, M. Hering and C. Vinzant, An algebraic characterization of injectivity in phase retrieval, {\it Appl. Comput. Harm. Anal.,} 38(2015), 346-356.


\bibitem{7Ariano} G.M. d’Ariano,  P. Perinotti, and M. F. Sacchi, Informationally complete measurements and group representation, {\it Journal of Optics B: Quantum and Semiclassical Optics,}  6.6 (2004): S487.

\bibitem{Edidin} D. Edidin, Projections and Phase Retrieval, {\it Appl. Comput. Harmon. Anal.,} 42 (2017), no. 2, 350-359.

\bibitem{Eldar} Y.C. Eldar, N. Hammen and D. Mixon, Recent Advances in Phase Retrieval, {\it  IEEE Signal Processing Magazine},  September 2016, 158--162.

\bibitem{9Eldar} Y. C. Eldar, Yonina and G. David Forney, Optimal tight frames and quantum measurement, {\it IEEE Transactions on Information Theory,}  48.3 (2002): 599-610.

\bibitem{4Gupta}  V. Gupta, P.  Mandayam and  V.S. Sunder,  The functional analysis of quantum information theory, A Collection of Notes Based on Lectures by Gilles Pisier, K. R. Parthasarathy, Vern Paulsen and Andreas Winter, Lecture Notes in Physics 902, Springer, 2015.

\bibitem{8Han} D. Han, and T. Juste, Phase-retrievable operator-valued frames and representations of quantum channels,  {\it Linear Algebra and its Applications,}  579 (2019), 148-168.

\bibitem{13Han} D. Han, and  D. Larson,  Frames, bases and group representations,  {\it Memoirs of Amer.  Math.  Soc.,}  Vol 697 (2000)

\bibitem{14Han} D. Han,  D. Larson, B. Liu and R. Liu, Operator-Valued Measures, Dilations, and the Theory of Frames, {\it Memoirs  Amer. Math. Soc.,}   Vol. 229, No.  1075 (2014).

\bibitem{Han-1} D. Han, Q. Hu and R. Liu, Injective continuous frames and quantum detections, {\it Banach J. Math. Anal.,} 15 (1) (2021), 1-24.

\bibitem{Han-2} D. Han, T. Juste, Y. Li and W. Sun, Frame phase-retrievability and exact phase-retrievable frames, {\it  J. Fourier Anal. Appl., } 25(2019), 3154-3173.

\bibitem{Han-3} D. Han and T. Juste, Phase-retrievable operator-valued frames and representations of quantum channels, {\it  Lin. Alg. Appl.,} 579(2019): 148-168.

\bibitem{Han-4} D. Han, P.  Li, B. Meng and W. Tang, Operator valued frames and structured quantum channels {\it Science in China –Mathematics,}  R. Kadison’s 85th birthday special issue. 54(2011), 2361–2372.

\bibitem{Han-5} D. Han and K. Liu, Orbit injective covariant quantum channels, preprint 2022.

\bibitem{18Holevo} A. Holevo, The Choi–Jamiolkowski forms of quantum Gaussian channels, {\it Journal of Mathematical Physics,}  52.4 (2011): 042202.

\bibitem{Kaftal} V. Kaftal, D. Larson and S. Zhang, Operator-valued frames,  {\it Trans. Amer. Math. Soc.,}   361 (2009): 6349--6385.

\bibitem{11Li} L.  Li, Lan, et al., Phase retrievable projective representation frames for finite abelian groups, {\it Journal of Fourier Analysis and Applications,}  25.1 (2019): 86-100.

\bibitem{3Nielsen}  M. Nielsen and Isaac Chuang,  Quantum computation and quantum information: 10th Anniversary Edition Anniversary Edition, Cambridge University Press 2000.

\bibitem{17Paulsen} V. Paulsen and F.  Shultz, Complete positivity of the map from a basis to its dual basis, {\it Journal of Mathematical Physics} 54.7 (2013): 072201.

\bibitem{Paulsen} V. Paulsen, Completely Bounded Maps and Operator Algebras, Cambridge Studies in Advanced Mathematics, 78, Cambridge University Press, 2003

\bibitem{Poumai}   K. T. Poumai, S. K. Kaushik and  S. V. Djordjevic, Operator valued frames and applications to quantum channels, International Conference on Sampling Theory and Applications (SampTA), 2017.

\bibitem{Renes} J. Renes, Frames, Designs, and Spherical Codes in
Quantum Information Theory, Ph.D Dissertation, The University of New Mexico, 2004.

\bibitem{5Renes} J. Renes,  R. Blume-Kohout, A. J. Scott and C. Caves  Symmetric informationally complete quantum measurements,  {\it Journal of Mathematical Physics,}  45.6 (2004): 2171-2180.


\bibitem{1Shankar} R.  Shankar,  Principles of quantum mechanics. Springer Science And Business Media, 2012.

\bibitem{6Tumalka} R. Tumulka,  POVM (Positive Operator Value Measure), {\it Compendium of Quantum Physics,} (2009): 480-484.

\bibitem{Wang-Xu-1} Y. Wang and Z. Xu, Generalized phase retrieval : measurement number, matrix recovery and beyond, {\it Applied and Computational Harmonic Analysis,} 47(2019, 423-446.

\bibitem{Wang-Xu-2}  Y. Wang and Z. Xu,  Phase retrieval for sparse signals, {\it Applied Computational Harmonic Analysis, } 37(2014), 531--544.

\bibitem{2Wilde} M. Wilde,  Quantum information theory. Cambridge University Press, 2013.





\end{thebibliography}
\end{document}